\pdfoutput=1
\documentclass[
    twocolumn,amsmath,amssymb
]{revtex4-1}

\usepackage{graphicx}
\usepackage{dcolumn}
\usepackage{bm}

\usepackage[utf8]{inputenc}
\usepackage[T1]{fontenc}
\usepackage{etoolbox}

\usepackage{algorithm}
\usepackage{algorithmic}

\usepackage[hidelinks]{hyperref}

\usepackage[capitalise]{cleveref}

\usepackage{multirow}

\usepackage{xcolor}

\setcitestyle{super}

\begin{document}

\title{Solvating machine-learning interatomic potentials for solution-phase chemistry}

\author{Anton Z. Ni}
\author{Joonho Lee}
\email{joonholee@g.harvard.edu}
\affiliation{
    Department of Chemistry and Chemical Biology, Harvard University, Cambridge, MA
}

\date{\today}

\begin{abstract}
Most machine-learning interatomic potentials (MLIPs) are trained on gas-phase data, which limits their use for chemistry in solution. Here, we introduce AnySolv, a transferable implicit solvation potential that provides solvent-dependent energy and force corrections to existing gas-phase potentials without retraining them. It is trained on 1.4 million solvation energies and forces computed with density functional theory (DFT) and SMD in 36 solvents, spanning charged, open-shell, and transition-state structures. Because solvents are represented by continuous physical descriptors, AnySolv supports 179 solvents, including those absent from training. On external benchmarks, AnySolv reproduces DFT-SMD solvation energies to 2--4~meV mean absolute error at more than three orders of magnitude lower cost, and it matches the accuracy of DFT-SMD against experimental solvation free energies. This enables screening of the mechanisms of 120 nucleophilic aromatic substitution reactions in six phases as well as molecular dynamics simulations that reproduce the experimentally observed solvent-dependent conformational preferences of alanine dipeptide. By providing transferable solvation corrections to existing MLIPs, AnySolv enables scalable simulations of solvent-dependent chemical reactivity and molecular dynamics.
\end{abstract}

\maketitle

\section{Introduction}
\label{sec:intro}

\begin{figure*}[t]
    \centering
    \includegraphics[width=\textwidth]{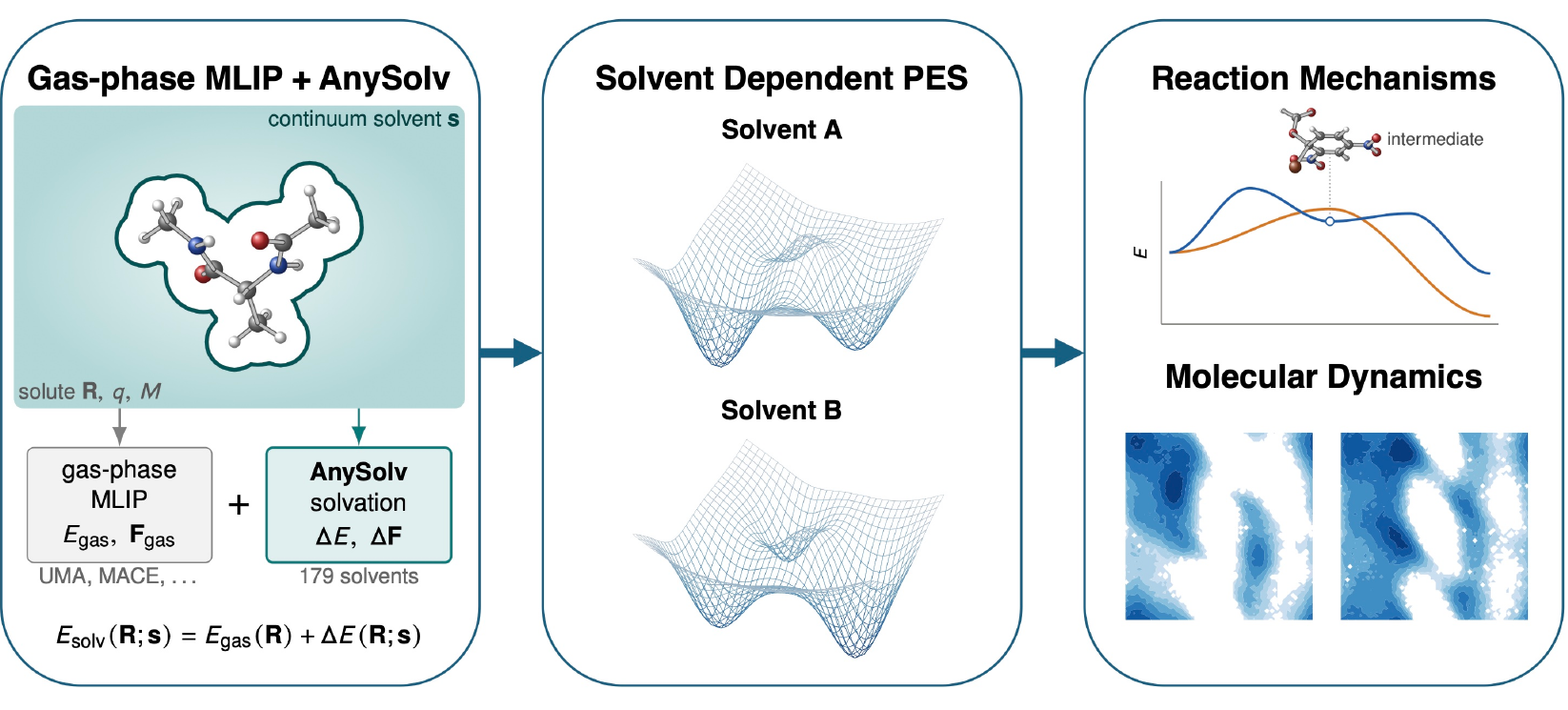}
    \caption{\textbf{Overview of AnySolv.} AnySolv predicts the solvation energy $\Delta E$ and forces $\Delta\mathbf{F}$ of a solute with coordinates $\mathbf{R}$, charge $q$, and spin multiplicity $M$ in a continuum solvent described by the descriptor vector $\mathbf{s}$, and adding them to any gas-phase MLIP gives the energy and forces in solution. From there, we can obtain solvent-dependent potential energy surfaces. This enables us to efficiently probe reaction mechanisms and perform molecular dynamics with implicit solvent.}
    \label{fig:concept}
\end{figure*}

Solvents can fundamentally alter chemical reactivity by stabilizing charged intermediates, shifting reaction barriers, and changing conformational equilibria.\cite{cramer199908,castejon199903}
Accurately capturing these effects is therefore essential for predictive simulations of solution-phase chemistry.
Machine-learning interatomic potentials (MLIPs) have recently emerged as computationally efficient alternatives to density functional theory (DFT), enabling atomistic simulations at scales previously inaccessible to electronic structure methods.\cite{batzner202205,batatia202212,anstine202504,fu202502,wood202506}
However, most general-purpose MLIPs are trained primarily on gas-phase reference data, limiting their applicability to solution-phase chemistry.\cite{levine202505}

One approach to incorporating solvent effects into MLIP simulations is to represent solvent molecules explicitly, often requiring hundreds to thousands of additional atoms.\cite{zhang202407,kovacs202505}
Alternatively, implicit solvation models describe the solvent through an effective potential, allowing geometry optimizations, transition-state searches, and molecular dynamics without explicitly representing solvent molecules.
Established approaches, including polarizable continuum models (PCM) and conductor-like screening models (COSMO), are widely used in electronic structure calculations.\cite{miertus198102,marenich200905,klamt199305,cramer199908}
In particular, SMD represents solvent effects through a set of macroscopic physical descriptors, allowing a single parametrization to describe a broad range of solvents.\cite{marenich200905,mnsddb,abraham1993}
However, the solvation potential in conventional DFT-based implicit solvation models is solved self-consistently with the electron density of the solute, which standard MLIPs do not provide.
Consequently, these models cannot be directly incorporated into MLIP simulations, and inexpensive semiempirical alternatives such as GFN2-xTB with ALPB generally offer lower accuracy.\cite{ehlert202107,bannwarth201903}

Several machine-learning approaches have sought to overcome these limitations, as recently reviewed in Ref.~\citenum{banchode202601}.
Some predict solvation free energies but cannot provide the nuclear forces needed for geometry optimization or molecular dynamics.\cite{lim201909,vermeire202108,chung202202}
Others provide forces but are trained on classical force fields, inheriting the limitations of their reference models.\cite{chen202108,katzberger202305,airas202311,katzberger202508}
Zubatyuk \emph{et al.} learned SMD-water energies and forces with AIMNet, but the model must be retrained for each solvent and base potential.\cite{zubatyuk201908}
More recently, ConSolv introduced solvent-conditioned learning that generalizes to solvents absent from training, but it is restricted to neutral solutes and does not provide a correction that can be added to arbitrary gas-phase potentials.\cite{zhang202606}
To our knowledge, no existing model provides a multi-solvent correction, with forces, that can be added to any gas-phase potential and remains valid for charged and open-shell solutes and at geometries far from equilibrium, such as transition states.

Here, we introduce AnySolv, an equivariant message-passing graph neural network that learns solvation energy and force corrections independently of the gas-phase potential.
Rather than predicting the total energy in solution, AnySolv learns the difference between solvated and gas-phase DFT energies at a given molecular geometry, together with the corresponding forces obtained as the negative energy gradient.
These corrections can be added to existing MLIPs without retraining.
We represent solvents with continuous physical descriptors,\cite{mnsddb} enabling a single model to support 179 solvents, including those absent from training.
We train AnySolv on approximately 1.4 million DFT-SMD calculations in 36 solvents at the $\omega$B97M-V/def2-TZVPD level of theory, using geometries drawn primarily from OMol25\cite{levine202505} that span charged and open-shell species as well as nonequilibrium structures along reaction pathways.

We demonstrate that AnySolv reproduces DFT-SMD solvation energies and forces, including for solvents absent from training, while retaining the accuracy of the reference method against experimental solvation free energies. Its compatibility with different gas-phase MLIPs and computational cost more than three orders of magnitude lower than DFT-SMD make previously impractical simulations feasible. In particular, we systematically examine solvent-dependent reaction mechanisms across 120 nucleophilic aromatic substitution reactions and obtain conformational free-energy landscapes of alanine dipeptide that reproduce experimentally observed solvent effects. Together, these results establish a general strategy for incorporating transferable solvation corrections into existing MLIP workflows. See \cref{fig:concept} for an overview.

\section{Theory and Methods}
\label{sec:methods}

AnySolv is an equivariant message-passing graph neural network that predicts molecular solvation energy. As with most other MLIPs,\cite{batzner202205,batatia202212,anstine202504,wood202506} the model takes nuclear coordinates, total charge, and spin multiplicity as input.
It also takes the solvent, encoded through a set of descriptors from the Minnesota Solvent Descriptor Database.\cite{mnsddb} It returns a single-point solvation energy $\Delta E$ and its gradient (\cref{sec:delta}).

The model follows FAIRChem's UMA-S architecture,\cite{wood202506,fairchem} which is built on eSEN\cite{fu202502} and computes the messages between atoms with the edgewise SO(2) convolutions introduced in eSCN.\cite{passaro202307} The eSEN architecture was chosen because it was designed to generate smooth potential energy surfaces. Node features are spherical-harmonic tensors up to degree two, updated over four message-passing layers with 128 channels and a 6~\AA{} cutoff. We read out an energy from each atom's scalar channel, sum these atomic energies, and obtain forces by differentiating the total energy with respect to the nuclear coordinates. 

We train two versions of AnySolv on the same training data. The first is fine-tuned from the UMA-S-1.2 checkpoint, a 291.3~M-parameter mixture-of-experts (MoE) model with 64 experts, after a solvent embedding is incorporated into its global feature vector. We call it the MoE model. The second, which we call the compact model, is a dense 6.4~M-parameter model trained from scratch, with the same layer structure but a single weight matrix in place of each expert mixture. Because the expert coefficients depend only on system-level inputs and are fixed for a given system, the MoE model can be merged into a dense network of the same size as the compact model at inference time, so the two models have similar inference costs (\cref{sec:inference}). Apart from the solvent embedding, all architectural details follow UMA-S-1.2 (\cref{app:arch}).

\subsection{Learning Solvation Energies}
\label{sec:delta}

Our target quantities are the single-point solvation energy and the corresponding forces:
\begin{gather}
    \Delta E(\mathbf{R}, q, M, \mathbf{s}) = E_\text{solv}(\mathbf{R}, q, M, \mathbf{s}) - E_\text{gas}(\mathbf{R}, q, M),
    \label{eq:delta} \\
    \Delta\mathbf{F} = -\nabla_{\mathbf{R}}\,\Delta E,
    \label{eq:delta-force}
\end{gather}
where $E_\text{solv}$ and $E_\text{gas}$ are the energies of the solute in solution and in vacuum, respectively, $\mathbf{R}$ denotes the nuclear coordinates, $q$ is the total charge, $M$ is the spin multiplicity, and $\mathbf{s}$ is the solvent descriptor vector (see \cref{sec:solvent}). 
These quantities form a single-point correction designed to be added to a standard computational workflow using MLIPs or quantum chemistry calculations, without explicit sampling of solvent configurations.
Experimental solvation free energies are available only for equilibrium ensembles and cannot provide labels at arbitrary geometries, so we generate labels with an implicit solvation model, SMD.\cite{marenich200905}
Its energy is a continuous function of solute geometry with analytic gradients, so we can generate consistent energy and force labels at any geometry. Many semiempirical and machine-learning implicit solvation models are instead parametrized using COSMO-RS,\cite{klamt199502,ehlert202107,vermeire202108,chung202401} for which analytic forces are not readily available, because the COSMO-RS free energy is evaluated in a one-shot statistical-thermodynamic step after the SCF. \Cref{sec:data} describes the label generation and training procedure.

\subsection{Solvent Embedding}
\label{sec:solvent}

In AnySolv, the solvent is represented by seven physical descriptors from the Minnesota Solvent Descriptor Database,\cite{mnsddb} as in SMD.\cite{marenich200905}
The solvent descriptor vector, $\mathbf s$, is then
\begin{equation}
    \mathbf{s} = \left(n,\ \alpha,\ \beta,\ \gamma,\ \varepsilon,\ \phi,\ \psi\right),
    \label{eq:solvent-vec}
\end{equation}
where $n$ is the index of refraction, $\alpha$ and $\beta$ are Abraham's hydrogen-bond acidity and basicity,\cite{abraham1993} $\gamma$ is the macroscopic surface tension, $\varepsilon$ is the dielectric constant, and $\phi$ and $\psi$ are the aromaticity and electronegative halogenicity (\cref{tab:descriptors}).
We use physical descriptors rather than other representations, such as graph-based representations or one-hot encodings of the solvent, because the descriptor space is continuous and allows interpolation between chemically related solvents.
The ConSolv authors independently reached a similar conclusion.\cite{zhang202606}
We show that our descriptor-based encoding generalizes well to solvents held out of training.

\begin{table}
    \centering
    \begin{tabular}{llcc}
    \hline\hline
    \noalign{\vskip 0.3em}
    Descriptor & Definition & $T_k$ & $\sigma_k$ \\[0.3em]
    \hline
    \noalign{\vskip 0.3em}
    $n$            & index of refraction (293~K)                 & $n-1$            & 0.0684 \\
    $\alpha$       & Abraham H-bond acidity, $\Sigma\alpha_2^H$  & --               & 0.182 \\
    $\beta$        & Abraham H-bond basicity, $\Sigma\beta_2^H$  & --               & 0.235 \\
    $\gamma$       & surface tension (cal\,mol$^{-1}$\,\AA$^{-2}$) & --             & 11.69 \\
    $\varepsilon$  & dielectric constant (298~K)                 & $1-1/\varepsilon$ & 0.165 \\
    $\phi$         & aromaticity                                 & --               & 0.322 \\
    $\psi$         & electronegative halogenicity                & --               & 0.175 \\[0.3em]
    \hline\hline
    \end{tabular}
    \caption{\textbf{Details of solvent descriptors.} We list the solvent descriptors, the transformation $T_k$ applied to each (\cref{eq:solvent-norm}; ``--'' denotes the identity), and the population standard deviation $\sigma_k$ of the transformed values over the 179 solvents of the Minnesota Solvent Descriptor Database. Aromaticity is the fraction of non-hydrogen atoms that are aromatic carbon; electronegative halogenicity is the fraction of non-hydrogen atoms that are F, Cl, or Br.}
    \label{tab:descriptors}
\end{table}

We transform the index of refraction $n$ and the dielectric constant $\varepsilon$. The index of refraction is shifted by its vacuum value, $n \to n - 1$, and the dielectric constant is transformed to the Born factor, $\varepsilon \to 1 - 1/\varepsilon$, which gives the dependence of the Born electrostatic solvation energy on the dielectric constant. This choice is motivated by the generalized Born model\cite{still199008} and ALPB with GFN2-xTB,\cite{ehlert202107} both of which have a similar dependence of the solvation energy on the dielectric constant.
Each transformed descriptor is then divided by its population standard deviation $\sigma_k$ over the 179 solvents in the database, without mean subtraction, so that the gas phase remains the zero solvent descriptor vector:
\begin{equation}
    \tilde{s}_k = \frac{T_k(s_k)}{\sigma_k},
    \qquad
    T_k(s_k) =
    \begin{cases}
        s_k - 1     & s_k = n, \\
        1 - 1/s_k   & s_k = \varepsilon, \\
        s_k         & \text{otherwise}.
    \end{cases}
    \label{eq:solvent-norm}
\end{equation}
The normalized vector $\tilde{\mathbf{s}} \in \mathbb{R}^{7}$ is mapped to a $C = 128$-dimensional solvent embedding by a two-layer perceptron with hidden width $h_s = 16$,
\begin{equation}
    \mathbf{e}_s = \mathbf{W}_2\,\operatorname{SiLU}\!\left(\mathbf{W}_1\,\tilde{\mathbf{s}} + \mathbf{b}_1\right) + \mathbf{b}_2 .
    \label{eq:solvent-mlp}
\end{equation}
We choose $C$ so that the solvent embedding has the same dimension as the other global features of UMA-S.
UMA-S conditions on the global graph scalars by mixing learned embeddings of the total charge, $\mathbf{e}_q$, the spin multiplicity, $\mathbf{e}_M$, and the source dataset, $\mathbf{e}_d$, into a single per-system vector $\mathbf{h}_\text{sys}$.\cite{wood202506}
We append the solvent embedding to these embeddings before mixing and leave everything downstream unchanged:
\begin{equation}
    \mathbf{h}_\text{sys} = \operatorname{SiLU}\!\left(\mathbf{W}_\text{mix}\,
        \left[\mathbf{e}_q;\ \mathbf{e}_M;\ \mathbf{e}_d;\ \mathbf{e}_s\right] + \mathbf{b}_\text{mix}\right),
    \label{eq:hsys-solvent}
\end{equation}
where $\mathbf{W}_\text{mix} \in \mathbb{R}^{C \times 4C}$.
The dataset embedding $\mathbf{e}_d$ is retained for compatibility with the pretrained weights but is constant here, since all training data carry a single dataset tag.

\begin{figure}
    \centering
    \includegraphics[width=\linewidth]{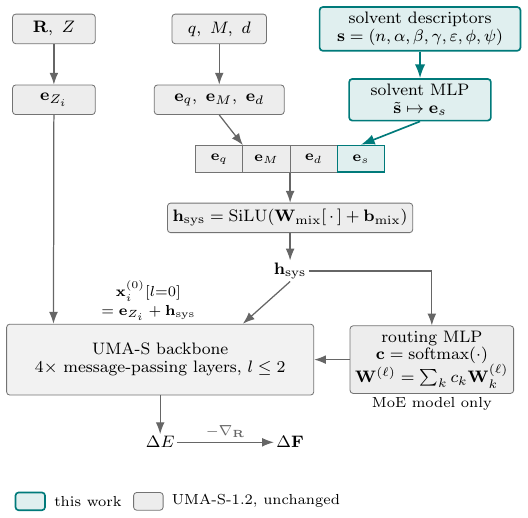}
    \caption{\textbf{Solvent embedding in AnySolv.} Teal blocks are introduced in this work; gray blocks are unchanged from UMA-S-1.2. A similar grafting can be applied to other graph neural network architectures.}
    \label{fig:conditioning}
\end{figure}

The vector $\mathbf{h}_\text{sys}$ is added to the scalar channel of each atom's node features at the input and at the entrance of every message-passing layer, and it also determines the coefficients that mix the expert weights in the MoE model. 
The experts are the weights of the SO(2) convolutions that compute the messages, so in the MoE model the solvent embedding modulates the message weights themselves. In contrast, the compact model, which has no experts, is conditioned on the solvent only through the node features (\cref{fig:conditioning}).

When the solvent block was grafted onto UMA-S-1.2, the pretrained $\mathbf{W}_\text{mix} \in \mathbb{R}^{C \times 3C}$ was zero-padded to a shape of $C\times 4C$, and the output layer of the solvent perceptron, \cref{eq:solvent-mlp}, was initialized near zero.
The weights of our message-passing backbone were initialized to the weights of UMA-S-1.2. The output perceptron of UMA-S-1.2, which maps each atom's features to its energy, predicts total energies rather than solvation energies, so we instead initialized it with random weights. The solvation energy depends on the same features of the solute that this backbone has learned to represent, such as its charge distribution and polarity, so we expect the pretrained backbone to be a better starting point than random weights. Fine-tuning also avoids training a 291~M-parameter model from scratch.
We initialized the compact model randomly at the start of its preliminary training (\cref{sec:data}).

\subsection{Datasets and training}
\label{sec:data}
Training labels (1.4~M in total) were obtained from SMD\cite{marenich200905} energy and force calculations with the $\omega$B97M-V range-separated hybrid exchange--correlation functional\cite{mardirossian201606} and the def2-TZVPD basis set\cite{weigend200508,rappoport201010} in ORCA 6.0.1.\cite{orca} The ORCA input string is provided in \cref{app:labels}. 
We took most solute geometries from the OMol25 dataset,\cite{levine202505} which stores gas-phase energies and forces at the same level of theory, so we performed only the solvated calculations. 

The labels store the differences in energy and nuclear gradients between the solvated and gas-phase calculations. Most labels come from reaction pathways and transition states. About 19\% of the conformers carry a net charge (from $-4$ to $+4$), and about 5\% are open-shell. A total of 36 solvents were sampled, chosen to broadly cover the descriptor space. To reduce the computational cost of generating labels, we restricted solutes in the 35 solvents other than water to at most 50 atoms. Water, which had no size cap, was assigned a larger share of the OMol25 conformers than the other solvents and has roughly five times as many labels as a typical solvent; 11\% of its labels come from solutes with more than 50 atoms. The solute sampling and the composition of the training set are described in \cref{app:solutes} and \cref{fig:distributions}.

Three smaller datasets fill gaps in the OMol25 data. First, to balance the charge distribution of the training set, we labeled 12,783 anion geometries (charges $-1$ and $-2$) from the AIMNet2 dataset\cite{anstine202504} at the same level of theory in acetonitrile, DMSO, methanol, and water, giving 46,166 training labels. Second, the training mix was augmented with 9,316 DFT-SMD labels ($\omega$B97M-V/def2-TZVPD, four solvents spanning Born factors 0.47--0.99) for 447 synthetic charge-separated species: $4\times4$ combinations of cationic and anionic groups on alkyl, oligoether, aryl--alkyl, and partially unsaturated linkers with charge separations up to 30~\AA, plus some charged control species and contact ion pairs, together upsampled to about 2.7\% of each training epoch. These species were included to improve coverage of systems with widely separated charges. The $\omega$-amino acids used in the long-range evaluation (\cref{sec:longrange}) were excluded by construction, and one group combination (pyridinium--phenolate) and one linker class (partially unsaturated) were held out from training to test generalization (\cref{app:augmentation}). 
Lastly, 211 experimental solvation free energies of monatomic solutes were included: 202 noble-gas values in 46 solvents from the Solv@TUM database\cite{hille201901} and 9 monatomic ions from MNSol,\cite{mnsol} together spanning 49 solvents. These are the only single atoms in the training set, since OMol25 contains no isolated atoms.
Because SMD is parametrized to reproduce experimental solvation free energies and a single atom has no internal degrees of freedom, the experimental values are directly comparable to the single-point correction for these solutes. All single-atom labels were assigned to the training set.

We split the total data 95/2.5/2.5 into training, validation, and test sets so that a given conformer appears in exactly one split across all solvents for which it was labeled. 
To guard against failed or improperly converged SCF solutions, labels with $|\Delta E| > 15$~eV or a force component above 50~eV/\AA{} were discarded (97 labels in total).
The final training set contains 1,355,281 labels, and per-solvent counts are given in \cref{app:data}.

We trained both models in two stages. 
First, a preliminary version of each model was trained on 5.46~M GFN2-xTB/ALPB\cite{ehlert202107,bannwarth201903,xtb} solvation labels over 21 solvents, supplemented by DFT-SMD water labels (\cref{app:hyper}), starting from UMA-S-1.2 with the grafted solvent embedding (\cref{sec:solvent}) for the MoE model and from random weights for the compact model. 
Second, we trained the models from these weights on the DFT-SMD labels. The released models come from the last run of this stage and were trained on the full DFT-SMD set described above, without ALPB labels. 

The loss is a weighted sum of the per-atom mean absolute error of $\Delta E$ and the L2 norm of the force error, with weights 20 and 2 following the default UMA fine-tuning recipe,\cite{fairchem} and with both targets normalized by a single scale factor, the root-mean-square deviation of the training labels (0.193~eV).
We used AdamW\cite{loshchilov201905} with a weight decay of $10^{-3}$, a 0.1-epoch linear warm-up followed by cosine decay to 1\% of the peak learning rate, and gradient-norm clipping at 100, at a global batch size of 16 molecules; the final run of each model used a peak learning rate of $5 \times 10^{-5}$ for 4 epochs (345,852 steps). Hyperparameters of each stage are listed in \cref{app:hyper}.
All parameters, including those of the pretrained backbone, were trainable.

\subsection{Inference}
\label{sec:inference}

In UMA-S, each linear layer of the SO(2) convolutions is a mixture of $K = 64$ experts,\cite{wood202506} $\mathbf{W} = \sum_{k} c_k\,\mathbf{W}_k$, whose coefficients are determined by a router from the embedding of the elemental composition and $\mathbf{h}_\text{sys}$ (\cref{eq:hsys-solvent}). The coefficients depend on the elemental composition, charge, spin, and solvent but not on $\mathbf{R}$, so at inference time each mixture can be contracted into a single dense matrix once and reused for repeated calculations on the same system and solvent. After contraction, the MoE model differs from the compact model only in the width of its radial basis (\cref{app:arch}) and has a similar inference cost. The MoE coefficients are recomputed only when the elemental composition, charge, spin, or solvent changes.

\section{Results and Discussion}
\label{sec:results}
We evaluate AnySolv against DFT-SMD and experiment on standard benchmark sets (MNSol, Solv@TUM, and FlexiSol) and on SAMPL6 partition coefficients, and we then apply it to charge-separated zwitterions, solvent effects in the Menshutkin reaction, an S$_\mathrm{N}$Ar mechanism screen, and free-energy sampling of alanine dipeptide by molecular dynamics.

\begin{table}
    \centering
    \footnotesize
    \setlength{\tabcolsep}{3pt}%
    \begin{tabular}{lrrrrrrr}
    \hline\hline
    \noalign{\vskip 0.3em}
    & & \multicolumn{3}{c}{MoE} & \multicolumn{3}{c}{compact} \\
    Solutes & $n$ & $\Delta E$ & $|\Delta\mathbf{F}|$ & $\cos$ & $\Delta E$ & $|\Delta\mathbf{F}|$ & $\cos$ \\[0.3em]
    \hline
    \noalign{\vskip 0.3em}
    all      & 36,847 & 0.016 & 0.026 & 0.970 & 0.020 & 0.030 & 0.963 \\
    neutral  & 27,523 & 0.012 & 0.021 & 0.974 & 0.014 & 0.024 & 0.970 \\
    cations  &  4,894 & 0.024 & 0.038 & 0.948 & 0.034 & 0.046 & 0.932 \\
    anions   &  4,430 & 0.028 & 0.038 & 0.971 & 0.042 & 0.052 & 0.953 \\[0.3em]
    \hline\hline
    \end{tabular}
    \caption{\textbf{Errors in solvation energy on the held-out test set.} We provide the solvation-energy MAE $\Delta E$ in eV, the force-magnitude MAE $|\Delta\mathbf{F}|$ in eV/\AA, and the mean cosine similarity between predicted and reference solvation forces.}
    \label{tab:testset}
\end{table}

\begin{table*}
    \centering
    \footnotesize
    \setlength{\tabcolsep}{3pt}%
    \begin{tabular}{lrrrrrrrr}
    \hline\hline
    \noalign{\vskip 0.3em}
    & & \multicolumn{3}{c}{MoE} & \multicolumn{3}{c}{compact} & DFT-SMD \\
    Solutes & $n$ & $\Delta E_\mathrm{sp}$ & $\Delta E_\mathrm{opt}$ & $\Delta G_\mathrm{solv}$ & $\Delta E_\mathrm{sp}$ & $\Delta E_\mathrm{opt}$ & $\Delta G_\mathrm{solv}$ & $\Delta E_\mathrm{sp}$ \\[0.3em]
    \hline
    \noalign{\vskip 0.3em}
    Solv@TUM neutral  & 4,510 & 0.92 ($-$0.27) & 0.89 ($-$0.24) & 0.99 ($-$0.48) & 0.92 ($-$0.26) & 0.89 ($-$0.22) & 0.98 ($-$0.45) & 0.93 ($-$0.24) \\
    FlexiSol neutral  &   529 & 2.48 ($+$1.29) & 2.46 ($+$1.03) & 2.47 ($+$0.52) & 2.48 ($+$1.29) & 2.45 ($+$1.03) & 2.45 ($+$0.57) & 2.51 ($+$1.34) \\
    MNSol neutral     & 2,530 & 0.83 ($-$0.24) & 0.84 ($-$0.35) & 0.95 ($-$0.52) & 0.82 ($-$0.22) & 0.83 ($-$0.33) & 0.93 ($-$0.50) & 0.83 ($-$0.22) \\
    MNSol cations     &   132 & 4.77 ($+$3.92) & 4.60 ($+$3.64) & 4.56 ($+$3.64) & 4.78 ($+$3.92) & 4.60 ($+$3.64) & 4.60 ($+$3.69) & 4.79 ($+$3.94) \\
    MNSol anions      &   222 & 5.59 ($+$3.25) & 5.08 ($+$2.40) & 5.17 ($+$2.60) & 5.73 ($+$3.57) & 5.19 ($+$2.70) & 5.29 ($+$2.92) & 5.49 ($+$3.13) \\[0.3em]
    \hline\hline
    \end{tabular}
    \caption{\textbf{Solvation free energies against experiment.} We provide the MAE and, in parentheses, the mean signed error (model minus experiment), both in kcal/mol, for the three protocols and both models, with our DFT-SMD single points at the level of theory of the training labels in the last column. The datasets are those of \cref{sec:smd}.}
    \label{tab:experiment}
\end{table*}

\subsection{Benchmark against DFT-SMD}
\label{sec:smd}
We first measured the performance of our models on the held-out test split of the training distribution: 36,847 labels spanning transition states, charged solutes, and open-shell species. 
The MoE and compact models achieve solvation-energy mean absolute errors (MAEs) of 0.016 and 0.020~eV, respectively, force-magnitude MAEs of 0.026 and 0.030~eV/\AA, and mean force cosine similarities of 0.97 and 0.96, with errors concentrated in charged solutes (\cref{tab:testset}).

We then tested AnySolv on three external solvation datasets, each with experimental reference values to which we return in \cref{sec:experiment}: MNSol\cite{mnsol} (2,884 solute--solvent pairs across 92 solvents, 354 of them involving ions, after removing the nine monatomic ions used as training labels), Solv@TUM\cite{hille201901} (4,510 neutral pairs across 87 solvents), and FlexiSol\cite{wittmann202512} (530 pairs of conformationally flexible solutes, including molecules with more than 50 atoms). 
Of these pairs, 32\% of MNSol (in 61 of its 92 solvents) and 45\% of Solv@TUM (in 56 of its 87 solvents) are in solvents absent from the training set, whereas all FlexiSol solvents are trained solvents.

For these datasets, we also performed DFT-SMD single-point calculations at every benchmark geometry at the same level of theory as the training labels and evaluated the models against these results. 
This separates the models' fidelity to their training labels from the accuracy of the reference method. 
The MoE model reproduces the DFT-SMD reference solvation energies to 0.0021~eV MAE on MNSol (ions included), 0.0027~eV on Solv@TUM, and 0.0043~eV on FlexiSol, and the solvation forces with magnitude MAEs of 0.003--0.004~eV/\AA{} and mean cosine similarities of 0.987--0.997.

Because solvation energies vary far more from solute to solute than, for a given solute, from solvent to solvent, it is possible for a model to score well on overall metrics without learning the solvent dependence.
For example, on Solv@TUM, a model that predicted each solute's mean solvation energy in every solvent would still reach a coefficient of determination of $R^2 = 0.88$ against the reference values across all solute--solvent pairs. 
We therefore need to test solvent discrimination directly. 

To do so, we take the Solv@TUM solutes with entries in at least four solvents, separately for the training solvents (171 solutes) and the unseen solvents (143 solutes). For each solute, we subtract its mean solvation energy across solvents from both the predicted and the DFT-SMD values. This isolates the solvent dependence we seek to analyze. For each of the two categories, we then compute $R^2$ of the predictions against the references over all solutes together, with respect to the identity line $y = x$.
Against DFT-SMD, the MoE model achieves $R^2 = 0.998$ on the 31 training solvents in Solv@TUM and 0.944 on the 56 unseen solvents. The compact model achieves an $R^2$ of 0.991 on training solvents and 0.898 on unseen solvents. If we score solute by solute, the median $R^2$ is 0.9996 on training and 0.986 on unseen solvents for the MoE model, and 0.9992 on training and 0.956 on unseen solvents for the compact model.

With UMA-S-1.2 as the gas-phase potential on one NVIDIA A100 MIG instance (3g.20gb), a single point in solution took 105~ms in total with the MoE model and 79~ms with the compact model, nearly independent of molecular size. The AnySolv correction accounted for 42 and 16~ms, respectively; the MoE model is slower here because it recomputes and contracts its expert mixtures whenever the elemental composition or solvent changes (\cref{sec:inference}). 
In contrast, DFT-SMD in ORCA 6.0.1 on 16 CPU cores (Intel Xeon Platinum 8480+, 2.0~GHz base and 3.8~GHz maximum turbo frequency) took 236~s on average on the held-out test set and 1,374~s on FlexiSol, whose molecules have up to 85 atoms. 
Therefore, our models reproduce both DFT-SMD solvation energies and their per-solute solvent dependence at a cost more than three orders of magnitude lower.

\subsection{Benchmark against experiments}
\label{sec:experiment}

For all subsequent geometry optimizations, we use UMA-S-1.2 as the gas-phase potential, following the protocol in \cref{app:bench}. For each of MNSol, Solv@TUM, and FlexiSol, we compare the single-point solvation energy $\Delta E_\mathrm{sp}$ at the dataset geometry, $\Delta E_\mathrm{opt}$ after geometry optimization of the gas-phase and solvated structures, and $\Delta G_\mathrm{solv}$, which adds harmonic vibrational free-energy corrections to both structures. The choice among these should matter little\cite{ribeiro201112} but varies between publications, so we report all three (\cref{tab:experiment}) and use $\Delta G_\mathrm{solv}$ wherever later sections need solvation free energies.

For neutral molecules, the MoE model reaches $\Delta G_\mathrm{solv}$ MAEs of 0.99~kcal/mol on Solv@TUM, 0.95~kcal/mol on MNSol, and 2.47~kcal/mol on FlexiSol. For all three neutral sets, the protocols agree with one another and with our DFT-SMD single points to within 0.12~kcal/mol, so the residual error reflects SMD accuracy rather than the model's representational power. Replacing UMA-S-1.2 with MACE-POLAR-1-M\cite{batatia202602} as the gas-phase potential changes the $\Delta E_\mathrm{opt}$ and $\Delta G_\mathrm{solv}$ MAEs by at most 0.05~kcal/mol in every row (\cref{tab:experiment-bases}). This showcases the transferability of AnySolv across different MLIPs without any retraining.

For the MNSol ions, the model and DFT-SMD are again very similar. The MoE $\Delta G_\mathrm{solv}$ MAE is 4.56~kcal/mol on the 132 cations and 5.17~kcal/mol on the 222 anions, against 4.79 and 5.49~kcal/mol for the DFT-SMD single points, with a positive bias of 2.4--3.9~kcal/mol throughout: SMD generally underestimates the solvation of ions, and the model, which reproduces DFT-SMD on these same ions to 0.16~kcal/mol, inherits this bias.

\subsection{Partition coefficients: SAMPL6 \texorpdfstring{$\log_{10} P$}{log10 P}}
\label{sec:logp}
We now test AnySolv on water/1-octanol partition coefficients in the SAMPL6 dataset, which contains 11 small molecules resembling fragments of small-molecule protein kinase inhibitors, with $\log_{10} P$ values measured potentiometrically.\cite{isik202001,isik201912} This test measures how well the models capture transfer free energies between two solvents. 

For each of the 11 compounds, we generated conformers with RDKit\cite{rdkit} and kept up to eight of the lowest-energy stable conformers (47 conformers in total). We then relaxed each conformer in water and 1-octanol using UMA-S-1.2 and AnySolv.
A full vibrational calculation for each relaxed conformer in both water and 1-octanol gave its free energy $G_{i,s}$; translational and rotational terms were omitted, as they change $\log_{10} P$ by less than 0.01. We then computed the conformational ensemble free energy in each solvent using the Boltzmann partition function, $G_s=-RT\ln\sum_i\exp(-G_{i,s}/RT)$. The water-to-octanol transfer free energy was evaluated either from these ensemble free energies or from the free energy of the lowest-energy conformer in each solvent, and $\log_{10} P$ follows from it.

The root-mean-square error (RMSE) against the experimental $\log_{10} P$ values is 0.51 (Boltzmann ensemble) and 0.53 (lowest-energy conformer) for the MoE model. The compact model gives nearly identical values (\cref{app:compact}).
This compares with RMSEs of 0.49--1.42 for the six DFT-SMD-based submissions to the blind SAMPL6 challenge and with 0.38 for the best physical method overall.\cite{isik202001} 
Both models therefore lie within the cluster of the best quantum-chemistry-based SAMPL6 submissions.

\begin{figure}[h]
    \centering
    \includegraphics[width=\linewidth]{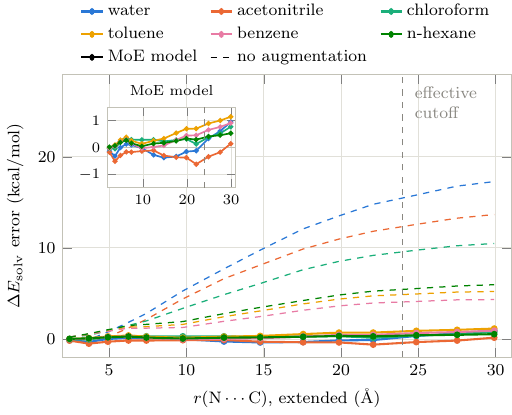}
    \caption{\textbf{Solvation energy errors for zwitterions.} Errors are shown for extended $\mathrm{H_3N^+}$--$(\mathrm{CH_2})_n$--$\mathrm{COO^-}$ zwitterions as a function of charge separation in six solvents, for the MoE model (solid) and a model trained without the charge-separation augmentation (dashed); the vertical line marks the 24~\AA{} effective cutoff. Inset: the MoE model errors on an expanded scale.
    }
    \label{fig:zwitterion}
\end{figure}

\subsection{Charge-separated zwitterions}
\label{sec:longrange}

We now test the ability of the models to learn long-range charge interactions, which probes the theoretical limit of the underlying architecture's representational power. With a radial cutoff of 6~\AA{} and four message-passing layers, the effective cutoff is 24~\AA{}. 
We therefore examined a series of $\omega$-amino-acid zwitterions, $\mathrm{H_3N^+}$--$(\mathrm{CH_2})_n$--$\mathrm{COO^-}$, from glycine ($n=1$) to the $n=23$ homolog, each in a fully extended linear conformation and in a cyclic conformation, in six solvents listed in order of increasing dielectric constant: $n$-hexane, benzene, toluene, chloroform, acetonitrile, and water. Single-point solvation energies were compared against DFT-SMD at the level of theory of the training labels, so any error due to the DFT functional is shared by the labels and the reference, and the comparison isolates the expressiveness of the architecture. 
To check for self-interaction error,\cite{cohen200808} which $\omega$B97M-V mitigates through its full long-range exact exchange, we computed the Hirshfeld charges of the ammonium and carboxylate groups in the gas-phase DFT calculations as the distance between them ranged from 10 to 30~\AA{}; they remained at $+0.66$ and $-0.84$, respectively. 
The series was excluded from the charge-separation data augmentation by construction, so only the shortest homologs could appear in the training data through the OMol25 subsets, and three of the six solvents (benzene, toluene, and acetonitrile) do not appear in the augmentation at all.

The results are summarized in \cref{fig:zwitterion}. For the extended chains, the mean absolute error of the MoE model in the single-point solvation energies across all $n$ and all six solvents is 0.013~eV (0.31~kcal/mol), with a maximum error of 0.049~eV (1.14~kcal/mol). For the cyclic structures, the mean absolute error is 0.008~eV (0.18~kcal/mol), with a maximum of 0.060~eV (1.39~kcal/mol). Beyond the effective cutoff, the extended-chain errors remain small: the mean absolute error is 0.026~eV (0.61~kcal/mol), and the cosine similarity between the predicted and reference solvation forces is at least 0.999 in every solvent. The compact model is comparably accurate on this series (\cref{app:compact}).

We also evaluated a model trained without the charge-separation augmentation (dashed lines in \cref{fig:zwitterion}). This model increasingly undersolvates the zwitterions as the chain length grows, with errors reaching 15.8~kcal/mol in water at 24.9~\AA{}. We therefore conclude that the augmented training labels are necessary for the model to learn longer-range charge interactions, and that they allow the model to capture these interactions up to and somewhat beyond its effective cutoff. The errors do not jump at 24~\AA{} because, even when the two charged termini are outside each other's environment, the mid-chain atoms remain within 24~\AA{} of both, so the network can still couple the two charges through them.

\begin{table*}[ht]
    \centering
    \footnotesize
    \setlength{\tabcolsep}{3.5pt}%
    \begin{tabular}{lrrrrrrrrrrrr}
    \hline\hline
    \noalign{\vskip 0.3em}
    & \multicolumn{3}{c}{$\Delta G^\ddagger$ (kcal/mol)} & \multicolumn{3}{c}{$\Delta G_\mathrm{rxn}$ (kcal/mol)} & \multicolumn{3}{c}{$r^\ddagger_\mathrm{C-Cl}$ (\AA)} & \multicolumn{3}{c}{$r^\mathrm{pdt}_\mathrm{C-Cl}$ (\AA)} \\
    Phase & DFT-SMD & MoE & compact & DFT-SMD & MoE & compact & DFT-SMD & MoE & compact & DFT-SMD & MoE & compact \\[0.3em]
    \hline
    \noalign{\vskip 0.3em}
    gas phase & 43.1 & 43.0 & 43.0 & 41.2 & 41.4 & 41.4 & 2.45 & 2.46 & 2.46 & 2.77 & 2.77 & 2.77 \\
    $n$-hexane & 32.8 & 32.7 & 33.5 & 18.9 & 18.8 & 18.9 & 2.33 & 2.32 & 2.32 & 3.00 & 3.06 & 3.04 \\
    chloroform & 27.0 & 26.8 & 28.2 & 1.1 & $-$0.3 & $-$5.5 & 2.27 & 2.26 & 2.24 & 3.25 & 3.43 & 4.60 \\
    methanol & 23.8 & 23.5 & 25.0 & $-$10.9 & $-$11.4 & $-$19.6 & 2.23 & 2.22 & 2.22 & 3.62 & 3.51 & 4.49 \\
    acetonitrile & 23.4 & 23.7 & 25.1 & $-$11.3 & $-$12.0 & $-$21.1 & 2.23 & 2.23 & 2.23 & 3.61 & 3.49 & 4.50 \\
    DMSO & 23.3 & 23.4 & 24.8 & $-$11.9 & $-$12.4 & $-$20.7 & 2.23 & 2.23 & 2.22 & 3.60 & 3.52 & 4.50 \\
    water & 23.1 & 24.0 & 25.2 & $-$11.8 & $-$12.7 & $-$21.2 & 2.23 & 2.22 & 2.22 & 3.54 & 3.59 & 4.57 \\[0.3em]
    \hline\hline
    \end{tabular}
    \caption{\textbf{Performance on the model Menshutkin reaction.} Activation free energies $\Delta G^\ddagger$ and reaction free energies $\Delta G_\mathrm{rxn}$ of $\mathrm{NH_3} + \mathrm{CH_3Cl} \to \mathrm{H_3NCH_3^+} + \mathrm{Cl^-}$, and the C--Cl distances at the transition state, $r^\ddagger_\mathrm{C-Cl}$, and in the product, $r^\mathrm{pdt}_\mathrm{C-Cl}$, from DFT-SMD and from UMA-S-1.2 with the MoE and compact AnySolv solvation energies.
    }
    \label{tab:menshutkin}
\end{table*}

\subsection{Solvent-dependent barriers: the Menshutkin reaction}
\label{sec:menshutkin}

The Menshutkin reaction,\cite{menshutkin1890} $\mathrm{NR_3} + \mathrm{R'X} \to \mathrm{R_3NR'^+} + \mathrm{X^-}$, separates charge along its S$_\mathrm{N}$2 reaction coordinate, so its barrier and reaction free energy depend strongly on the solvent.\cite{castejon199903,acevedo201007} We computed both for the model reaction $\mathrm{NH_3} + \mathrm{CH_3Cl} \to \mathrm{H_3NCH_3^+} + \mathrm{Cl^-}$ in the gas phase and in six solvents and compared them with DFT-SMD (\cref{tab:menshutkin}).\cite{gao199109}

Solvation lowers the DFT-SMD free-energy barrier from 43.1~kcal/mol in the gas phase to 23.1 in water and turns the reaction from endergonic ($+41.2$~kcal/mol) to exergonic ($-11.8$). The MoE model reproduces every solvated barrier to within 0.9~kcal/mol and the reaction free energies to within 1.4~kcal/mol, and its transition-state C--N and C--Cl distances agree with those of DFT-SMD to within 0.01~\AA{}. 
The compact model overestimates the free-energy barriers by 0.7--2.0~kcal/mol. 
In chloroform and polar solvents, its product relaxes to a more separated ion pair, whose solvation it overestimates (by 9.7~kcal/mol in water), making its reaction free energies 6.6--9.8~kcal/mol too negative. 
For comparison, substituting GFN2-xTB/ALPB\cite{ehlert202107} solvation energies for the AnySolv correction at the same MoE-relaxed geometries overestimates the solvated barriers by 1.2--6.0~kcal/mol and misses the reaction free energies by up to 9.6~kcal/mol in either direction.

The MoE model thus outperforms the compact model on the reaction barriers, and much more so on the reaction free energies. The compact model oversolvates the charge-separated ion pair: its solvation energies are similar to those of DFT-SMD for the reactant complex, where the charges have not yet separated, but deviate increasingly as the cation and anion separate. Thus, for reaction pathways in which charge is separated, we recommend the MoE model over the compact model.

Because AnySolv only adds a solvation correction, it should transfer to any gas-phase MLIP that accurately describes the gas-phase surface. To test this, we repeated the Menshutkin reaction calculations with the MoE model added to four other gas-phase MLIPs: MACE-POLAR-1-M\cite{batatia202602} and OrbMol-v2,\cite{rhodes202504} which were trained on OMol25, and AIMNet2\cite{anstine202504} and AIMNet2-NSE,\cite{kalita202601} which were trained on a different dataset at the $\omega$B97M-D3(BJ)/def2-TZVPP level of theory. The errors against DFT-SMD are tabulated in \cref{tab:menshutkin-bases}. The OMol25-trained models reproduce the gas-phase barrier and reaction free energy to within 0.5~kcal/mol, and with the MoE model they reproduce the solvated barriers and reaction free energies with mean absolute errors of 0.3--0.5 and 0.4--0.8~kcal/mol, respectively. 
AIMNet2 and AIMNet2-NSE overestimate the gas-phase barrier by 4.7 and 9.4~kcal/mol, respectively, so their solvated barriers are also too high in every solvent. In addition, for AIMNet2 in methanol, acetonitrile, and DMSO and for AIMNet2-NSE in $n$-hexane, relaxing the converged DFT-SMD product gave a structure in which Cl$^-$ is hydrogen-bonded to an N--H of the ammonium group rather than the expected backside ion pair.
Because of its large charge separation, the Menshutkin reaction is a good stress test of the transferability of our solvation model. Because AnySolv performs well with all suitable gas-phase potentials, we conclude that AnySolv transfers to gas-phase MLIPs that reproduce the gas-phase surface at the level of theory of our labels. When paired with other gas-phase models, AnySolv may inherit their errors.

\begin{table}
    \centering
    \footnotesize
    \setlength{\tabcolsep}{3pt}%
    \begin{tabular}{lrrrrrr}
    \hline\hline
    \noalign{\vskip 0.3em}
    & \multicolumn{2}{c}{Gas phase} & \multicolumn{4}{c}{Solvated} \\
    & & & \multicolumn{2}{c}{$\Delta G^\ddagger$} & \multicolumn{2}{c}{$\Delta G_\mathrm{rxn}$} \\
    Gas-phase model & $\Delta G^\ddagger$ & $\Delta G_\mathrm{rxn}$ & MSE & MAE & MSE & MAE \\[0.3em]
    \hline
    \noalign{\vskip 0.3em}
    \multicolumn{7}{l}{\textit{Trained on OMol25}} \\[0.15em]
    UMA-S-1.2 & 0.0 & $+$0.2 & $+$0.1 & 0.3 & $-$0.7 & 0.7 \\
    MACE-POLAR-1-M & $+$0.4 & $+$0.2 & $+$0.5 & 0.5 & $-$0.4 & 0.4 \\
    OrbMol-v2 & $+$0.2 & $+$0.1 & $+$0.2 & 0.4 & $-$0.8 & 0.8 \\
    \noalign{\vskip 0.3em}
    \hline
    \noalign{\vskip 0.3em}
    \multicolumn{7}{l}{\textit{Trained on the AIMNet2 dataset}} \\[0.15em]
    AIMNet2 & $+$4.7 & $+$5.0 & $+$6.5 & 6.5 & $+$3.8$^a$ & 3.8$^a$ \\
    AIMNet2-NSE & $+$9.4 & $+$1.5 & $+$10.1 & 10.1 & $+$1.3$^a$ & 1.3$^a$ \\[0.3em]
    \hline\hline
    \end{tabular}
    \caption{\textbf{Transferability of AnySolv across different MLIPs.} Errors against DFT-SMD (kcal/mol) of the activation free energies $\Delta G^\ddagger$ and reaction free energies $\Delta G_\mathrm{rxn}$ of $\mathrm{NH_3} + \mathrm{CH_3Cl} \to \mathrm{H_3NCH_3^+} + \mathrm{Cl^-}$ with the MoE model added to different gas-phase MLIPs. Gas phase: signed error of the gas-phase model alone, without AnySolv. Solvated: mean signed (MSE) and mean absolute (MAE) errors over the six solvents of \cref{tab:menshutkin}. $^a$Excluding solvents in which the product relaxes to a different ion pair (see text).}
    \label{tab:menshutkin-bases}
\end{table}

\begin{figure*}[ht]
    \centering
    \includegraphics[width=\textwidth]{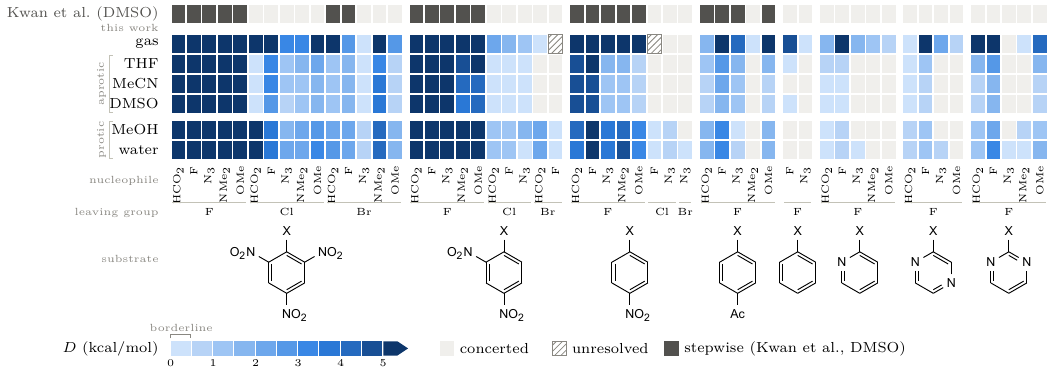}
    \caption{\textbf{S$_\mathrm{N}$Ar mechanism screening with UMA-S-1.2 and the MoE model.} Each column is one of the 54 reactions that have an intermediate in at least one phase, grouped by substrate, leaving group, and nucleophile, and each cell gives the electronic well depth $D$ of its Meisenheimer complex in one phase. The top row gives Kwan et al.'s B3LYP-D3(BJ)/6-31+G*/PCM classification of the same reactions in DMSO.\cite{kwan201809} The first color bin, $0 < D \le 0.5$~kcal/mol, is the borderline class. The 66 reactions without an intermediate in any phase are not shown.}
    \label{fig:snar}
\end{figure*}

\subsection{\texorpdfstring{S\textsubscript{N}Ar}{SNAr} Mechanism Screen}
\label{sec:snar}

A fast, reliable solvation model combined with state-of-the-art MLIPs makes it practical to screen reaction mechanisms across many substrates and solvents. We demonstrate this with nucleophilic aromatic substitution (S$_\mathrm{N}$Ar),\cite{rohrbach201911} long described as a two-step addition--elimination through an anionic intermediate, the Meisenheimer complex. Kwan et al.\ showed with kinetic isotope effects and a computational survey of 120 reactions that most S$_\mathrm{N}$Ar reactions are instead concerted, and that they are predicted to be stepwise mainly when a strongly electron-withdrawing ring is combined with fluoride as the nucleophile or the leaving group.\cite{kwan201809} Because the Meisenheimer well, where it exists, is only a few kcal/mol deep, the mechanism can change with the substrate, the nucleophile, and the solvent.
We took the 120 S$_\mathrm{N}$Ar reactions of Kwan et al.\ and located their stationary points with UMA-S-1.2 and the AnySolv MoE model in the gas phase and in THF, acetonitrile, DMSO, methanol, and water, connecting every transition state to its minima by an intrinsic reaction coordinate calculation (\cref{app:bench}). Although S$_\mathrm{N}$Ar reactions are often run in DMSO, many can also be run in other polar solvents. 

We classified each reaction by the electronic well depth $D$ of the Meisenheimer complex. The complex lies between two transition states, one for addition of the nucleophile and one for loss of the leaving group, and $D$ is the energy of the lower of the two relative to the complex. We call a reaction stepwise if $D > 0.5$~kcal/mol, borderline if $0 < D \le 0.5$~kcal/mol, and concerted if the Meisenheimer complex is not a minimum or no barrier separates it from the reactants or products (\cref{fig:snar}). We compare our screen with the classification of Kwan et al., obtained at the B3LYP-D3(BJ)/6-31+G*/IEFPCM(DMSO) level of theory. Because the two calculations differ in the underlying functional, basis set, and continuum solvation model, we look for agreement in trends rather than in individual well depths.

A screen of this scale requires thousands of geometry optimizations and transition-state searches, each followed by a frequency calculation. For the 720 reaction--phase pairs, we ran about 2,100 minimizations and 1,900 transition-state searches and followed each transition state found with an intrinsic reaction coordinate calculation. The full screen took 68~GPU-hours on NVIDIA A100 MIG instances.

Our screen reproduces the trends of Kwan et al.'s work (\cref{fig:snar}). In DMSO, the model finds an intermediate for all 21 reactions that Kwan et al.\ classified as stepwise, and the two studies agree on whether an intermediate exists for 101 of the 120 reactions. In 14 of the 19 disputed cases, our wells are shallower than 1~kcal/mol. None of these 19 intermediates is created by the solvation correction: all are already minima on the gas-phase UMA-S-1.2 surface, and solvation in DMSO makes every one of them shallower. 

Consistent with their findings, reactions with fluoride as the leaving group are the most likely to form a stable Meisenheimer complex, and the deepest wells occur on rings activated by nitro groups. On rings with less stabilizing substituents, or with a better leaving group, the local minimum of the Meisenheimer complex becomes shallower or disappears entirely. 
Across solvents, the predicted mechanism barely changes in THF, acetonitrile, and DMSO, whereas the wells are about 1~kcal/mol deeper in methanol and water.
Thus, the model reproduces the trends of a DFT study at a small fraction of the cost of DFT. 
This makes AnySolv suitable for high-throughput screening of substrates and solvents at near-DFT quality.

\begin{figure*}[ht]
    \centering
    \includegraphics[width=\textwidth]{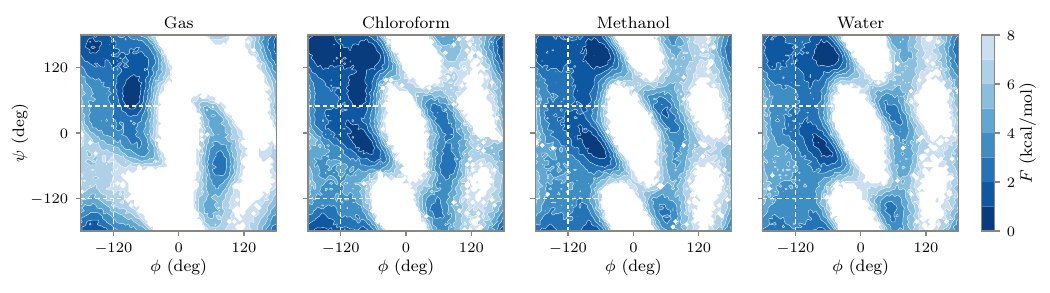}
    \caption{\textbf{Free-energy surfaces of alanine dipeptide.} We show the free-energy surfaces over the backbone dihedrals $\phi$ and $\psi$ from well-tempered metadynamics with UMA-S-1.2 and the MoE model at 298.15~K in the gas phase, chloroform, methanol, and water.}
    \label{fig:alanine}
\end{figure*}

\subsection{Conformational Ensembles from Molecular Dynamics}
\label{sec:alanine_md}

We now consider alanine dipeptide (Ace-Ala-NMe), a standard benchmark for solvation models because its conformations are described by the two backbone dihedrals $\phi$ (C--N--C$\alpha$--C) and $\psi$ (N--C$\alpha$--C--N).\cite{ramachandran196307,smith199909} The conformational distribution is also strongly solvent-dependent.\cite{madison198007} We sampled the free-energy surface of alanine dipeptide with UMA-S-1.2 and the MoE model by well-tempered metadynamics\cite{laio200210,barducci200801} on $\phi$ and $\psi$ in the gas phase and in chloroform, methanol, and water. 

Our training set contains 24 single-point labels for seven alanine dipeptide conformers; one of these labels is in methanol, and none are in chloroform or water. We ran the dynamics in ASE\cite{ase} with UMA-S-1.2\cite{wood202506,fairchem} and the MoE model, using a Langevin thermostat at 298.15~K and a 1~fs time step, and applied the bias with PLUMED,\cite{tribello201402} with eight walkers per phase sharing one bias\cite{raiteri200603} for 250~ps each (see \cref{app:bench} for more details). 
In a separate 50~ps microcanonical test (\cref{app:bench}), the total energy with AnySolv fluctuated within 1.3~meV/atom at a 1~fs time step and drifted by less than 0.002~meV/atom per ps, the same as with UMA-S-1.2 alone.

Alanine dipeptide has four main conformers in solution, which we name following Madison and Kopple.\cite{madison198007} In C7 ($\phi, \psi \approx -80^\circ, 80^\circ$), a hydrogen bond between the acetyl C=O and the N-methylamide N--H closes a seven-membered ring, as in a $\gamma$-turn; C5 ($-150^\circ, 150^\circ$) is the extended, $\beta$-strand-like conformer, closed only by a weak N--H$\cdots$O=C contact within the alanine residue; and PPII ($-80^\circ, 150^\circ$) and $\alpha_\mathrm{R}$ ($-80^\circ, -50^\circ$) have no intramolecular hydrogen bond, leave both amide groups exposed to the solvent, and have larger dipole moments. Chen et al.'s machine-learned implicit solvation model for alanine dipeptide, ISSNet, was trained on an explicit-water force-field simulation of that molecule,\cite{chen202108} so each new solvent would require a new reference simulation and a new model. 
By contrast, we can run AnySolv in all three solvents without any additional training or reference calculations.

\Cref{fig:alanine} shows the resulting free-energy surfaces. As the solvent becomes more polar, the equilibrium moves further away from C7 toward the PPII and $\alpha_\mathrm{R}$ basins. This is consistent with our expectations, as more polar solvents should solvate the N--H and C=O groups better. Madison and Kopple tracked this redistribution by circular dichroism and by the vicinal coupling constant $^3J(\mathrm{H^N,H^\alpha})$ between the amide proton and the hydrogen attached to the $\alpha$ carbon. We computed approximate coupling constants from the Karplus relation, averaged over the reweighted ensemble (\cref{app:bench}), and obtained $^3J(\mathrm{H^N,H^\alpha})$ values of 7.4~Hz in chloroform, 6.4~Hz in methanol, and 5.9~Hz in water, compared with measured values of 7.5, 6.8, and 6.0~Hz.\cite{madison198007} For comparison, we applied the same analysis to the solute configurations that Chen et al.\ distributed from their 1~$\mu$s explicit-solvent simulation with the AMBER ff99SB-ILDN force field\cite{lindorfflarsen201006} and TIP3P water,\cite{jorgensen198307} on which they trained their ISSNet models.\cite{chen202108} Relative to PPII, the basin free energies of that simulation differ from ours by at most 0.5~kcal/mol, but it gives $^3J(\mathrm{H^N,H^\alpha}) = 7.3$~Hz in water, an error of 1.3~Hz compared with 0.1~Hz for UMA-S-1.2 with AnySolv. This error arises because the ff99SB-ILDN backbone potential places the PPII and $\alpha_\mathrm{R}$ basins at more negative $\phi$ and puts about twice as much of the population in C5.

We also tested whether the ensembles generated by AnySolv match those of the DFT-SMD surface by recomputing 50 frames from each chloroform, methanol, and water simulation at the same level of theory as the training labels.
Because the populations depend only on energy differences between frames, we subtracted the mean over the frames of each solvent from both the AnySolv and the DFT-SMD solvation energies, analogous to the centering in \cref{sec:smd}. The centered energies agree with $R^2$ of 0.9996, 0.9997, and 0.9992 in chloroform, methanol, and water, respectively, and the solvation forces have mean cosine similarities above 0.9998.
By contrast, GFN2-xTB/ALPB\cite{ehlert202107} at the same frames reaches only $R^2 = 0.79$, 0.79, and 0.85 in chloroform, methanol, and water after centering, with errors of up to 0.12~eV (2.8~kcal/mol), several times $k_\mathrm{B}T$. 

Lastly, we note that, on average, one molecular dynamics step with UMA-S-1.2 and AnySolv took 84~ms on one A100 MIG instance, whereas a DFT-SMD energy and gradient for the same molecule took 182~s on 16 CPU cores, averaged over the 150 calculations above; at that cost, the two million force evaluations of each solvated simulation would have taken more than a decade. We thus conclude that AnySolv can be used as an implicit solvent term in molecular dynamics wherever {\it ab initio} molecular dynamics with SMD solvation is appropriate. Explicit-solvent simulations can capture effects that a continuum cannot, such as specific hydrogen bonds between solute and solvent. With a gas-phase MLIP, however, the same simulations would be much more expensive, if not prohibitive. Each solvent would require an equilibrated periodic box with several hundred solvent molecules, roughly 100 times as many atoms as the solute.

\section{Conclusions}
\label{sec:conclusions}

Machine-learning interatomic potentials have substantially expanded the scope of atomistic simulations, but their predominant reliance on gas-phase training data limits their applicability to chemistry in solution. Here, we have addressed this limitation by developing AnySolv, a transferable implicit solvation potential that can be combined with existing gas-phase models without retraining them. AnySolv provides solvation energies and forces for diverse solutes, including charged species and transition states, and, because it represents solvents through continuous physical descriptors, also in solvents absent from training. This modular approach separates the description of gas-phase chemistry from that of solvation, allowing a single model to introduce solvent effects into a wide range of computational workflows.

AnySolv reproduces DFT-SMD solvation energies and forces at more than three orders of magnitude lower computational cost, and it matches the accuracy of DFT-SMD against experimental solvation free energies. Crucially, this accuracy extends beyond equilibrium structures to solvent-dependent reaction energetics and conformational free-energy landscapes. A mechanistic screen of 120 nucleophilic aromatic substitution reactions in the gas phase and five solvents, comprising about 4,000 geometry optimizations and transition-state searches, took 68~GPU-hours and reproduced the trends of an earlier DFT survey, and metadynamics of alanine dipeptide reproduced experimental coupling constants in three solvents to within 0.4~Hz. Together, these results show that machine-learned solvation potentials can enable chemically meaningful simulations that would otherwise be computationally prohibitive at the DFT level.

The accuracy of AnySolv remains limited by the underlying SMD reference and the gas-phase potential with which it is combined. Nevertheless, our results establish a general strategy for incorporating DFT-SMD-quality implicit solvation into MLIPs without explicitly representing solvent molecules or developing solvent-specific models. By bridging the gap between gas-phase MLIPs and solution-phase chemistry, AnySolv opens a route toward routine, high-throughput investigations of solvent-dependent reactivity and molecular dynamics.

\section*{Code and Data Availability}
AnySolv is distributed as a standalone open-source Python package, \texttt{anysolv}, on GitHub; the model weights are hosted on HuggingFace, and the fork of FAIRChem used to build the solvent embedding is available as \texttt{fairchem-solvation}.\cite{anysolv} Calling \texttt{predict\_solvation\_energy} returns $\Delta E$ and $\Delta\mathbf{F}$, which are to be added to the energy and forces of any gas-phase method (MLIP or DFT). Both the compact and MoE models are publicly available.
The data behind the numbers in this paper are collected in a separate repository, \texttt{anysolv-publication-data}.\cite{anysolvdata}

\section*{AI Statement}
We used Anthropic's Opus 4.8 and Fable 5 to assist writing code for this project. We used Anthropic’s Fable 5.1 in the preparation of the manuscript. All uses were with supervision and verification by the authors.

\begin{acknowledgments}
This work was supported by the Air Force Office of Scientific Research Young Investigator Program under Award No. FA9550-26-1-B079. The work of A. Z. N. was partially supported by the NSF GRFP Fellowship for the initial stage of the work. We thank Corin Wagen and Meng-Fu Chen for insightful discussions during the project.
This work used computational resources from the FASRC cluster supported by the Faculty of Arts and Sciences (FAS) Division of Science Research Computing Group at Harvard and GPUs from RunPod.
\end{acknowledgments}

\bibliographystyle{naturemag}
\bibliography{refs}

\onecolumngrid
\clearpage
\appendix{}
\crefalias{section}{appendix}
\crefalias{subsection}{subappendix}
\renewcommand{\thefigure}{A\arabic{figure}}
\setcounter{figure}{0}
\renewcommand{\thetable}{A\arabic{table}}
\setcounter{table}{0}
\begin{center}
    \textbf{Appendix}
\end{center}

\section{Models and training}
\label{app:models}

\subsection{Architecture}
\label{app:arch}

Both models use the eSEN-based message-passing backbone of UMA-S with four layers, $C = 128$ spherical channels, $\ell_\text{max} = m_\text{max} = 2$, a 6~\AA{} radial cutoff with at most 300 neighbors per atom, 128 edge and hidden channels, gated activations, spectral feed-forward blocks, and RMS normalization of the spherical-harmonic features. Interatomic distances are expanded in Gaussians evenly spaced on $[0, 6]$~\AA, and the radial function of each SO(2) convolution is a two-hidden-layer perceptron on the concatenation of this expansion with learned embeddings of the source and target elements. Charge and spin multiplicity enter through learned tables, the energy is read out from the $\ell = 0$ channel of each atom by a three-layer perceptron and summed over atoms, and forces are the negative gradient of the energy. \Cref{tab:architecture} lists the two models side by side. Outside the expert mixture the only difference between them is the width of the radial basis, 32 Gaussians in the MoE model and 64 in the compact model. Of the 291.3~M parameters of the MoE model, 289.4~M are the 64-expert tensors of its 24 SO(2) linear layers (the $m = 0, 1, 2$ blocks of the two SO(2) convolutions in each layer) and 70~k form the router and the composition embedding from which the mixing coefficients are computed; merged for a single system it has 6,348,340 parameters, against 6,368,052 for the compact model.

\begin{table}[htb]
    \centering
    \footnotesize
    \begin{tabular}{lll}
    \hline\hline
    \noalign{\vskip 0.3em}
    & MoE & compact \\[0.3em]
    \hline
    \noalign{\vskip 0.3em}
    initialization & UMA-S-1.2 checkpoint, solvent block grafted & random \\
    parameters, total & 291,303,796 & 6,368,052 \\
    \quad expert SO(2) weights & 289,406,976 & -- \\
    \quad router and composition embedding & 70,464 & -- \\
    parameters after merging for one system & 6,348,340 & 6,368,052 \\
    experts per SO(2) linear layer & 64 & 1 \\
    SO(2) linear layers carrying experts & 24 (6 per layer) & 0 \\
    message-passing layers & 4 & 4 \\
    spherical channels $C$ & 128 & 128 \\
    $\ell_\text{max}$ / $m_\text{max}$ & 2 / 2 & 2 / 2 \\
    radial cutoff / max.\ neighbours & 6~\AA{} / 300 & 6~\AA{} / 300 \\
    edge / hidden channels & 128 / 128 & 128 / 128 \\[0.3em]
    \hline\hline
    \end{tabular}
    \caption{Architecture of the two models. Both share the eSEN-based backbone of UMA-S.}
    \label{tab:architecture}
\end{table}

\subsection{Training schedule and hyperparameters}
\label{app:hyper}

Each model was trained in two stages (\cref{tab:training}), with the individual runs listed in \cref{tab:runs}. The preliminary stage used inexpensive GFN2-xTB/ALPB labels\cite{bannwarth201903,ehlert202107} to test whether the solvation energy could be learned at all before DFT-SMD labels were generated. These labels were computed with GFN2-xTB at default settings as the difference between the ALPB and gas-phase single points at the same geometry, for 5.27~M OMol25 geometries in 21 solvents and 189~k AIMNet2 anion geometries, and were combined with the 173~k DFT-SMD water labels and the nine experimental anions from MNSol. The MoE model kept the UMA-S-1.2 backbone with a newly initialized energy head, since the pretrained head predicts total energies, and the compact model was initialized randomly. Both used the loss and optimizer of \cref{sec:data}. The final models were initialized from these weights on the assumption that the structure of the solvation energy learned at this stage carries over to DFT-SMD. The final stage then used the DFT-SMD labels of \cref{sec:data}.

\begin{table}[htb]
    \centering
    \footnotesize
    \begin{tabular}{lll}
    \hline\hline
    \noalign{\vskip 0.3em}
    & ALPB Pretraining & Training with SMD Labels \\[0.3em]
    \hline
    \noalign{\vskip 0.3em}
    initialization & MoE: UMA-S-1.2 backbone; compact: random & pretrained weights \\
    labels & GFN2-xTB/ALPB, 21 solvents (5.46~M) + DFT-SMD water (173~k) & DFT-SMD, 36 solvents (1.35~M) \\
    unique training labels & 5,633,695 & 1,355,281 \\
    labels sampled per epoch & 3,356,976 & 1,383,400 \\[0.3em]
    \hline\hline
    \end{tabular}
    \caption{Training data and settings of the preliminary GFN2-xTB/ALPB stage and the DFT-SMD stage. Epochs, peak learning rates, steps, and hardware of every run are given in \cref{tab:runs}; the loss, optimizer, and schedule are listed in \cref{sec:data}.}
    \label{tab:training}
\end{table}

\begin{table}[htb]
    \centering
    \footnotesize
    \begin{tabular}{lccccc}
    \hline\hline
    \noalign{\vskip 0.3em}
    stage & model & epochs & peak learning rate & steps & GPUs \\[0.3em]
    \hline
    \noalign{\vskip 0.3em}
    ALPB Pretrain & MoE & 4 & $1.5 \times 10^{-4}$ & 839,244 & $4 \times$ H200 \\
    ALPB Pretrain & compact & 8 & $4 \times 10^{-4}$ & 1,678,488 & $4 \times$ H200 \\
    DFT-SMD & MoE & 4 & $5 \times 10^{-5}$ & 345,852 & $4 \times$ H100 \\
    DFT-SMD & compact & 4 & $5 \times 10^{-5}$ & 345,852 & $2 \times$ A100 \\[0.3em]
    \hline\hline
    \end{tabular}
    \caption{Final training run of each stage.}
    \label{tab:runs}
\end{table}

\begin{figure}[htb]
    \centering
    \includegraphics[width=0.9\linewidth]{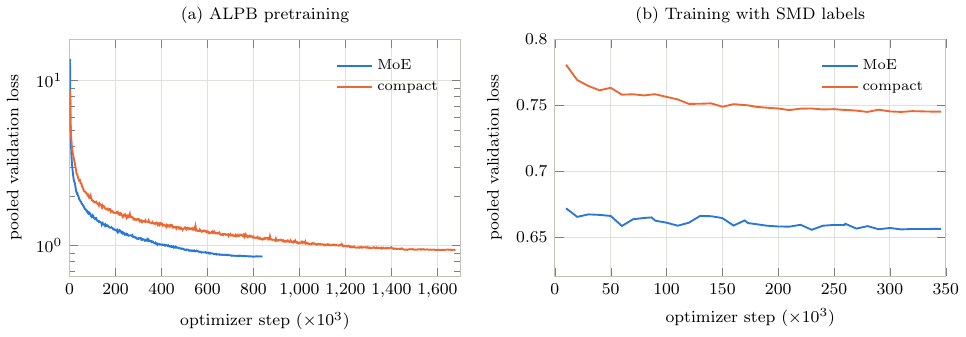}
    \caption{Pooled validation loss against optimizer step for the MoE (blue) and compact (orange) models during (a) ALPB pretraining and (b) training with SMD labels.}
    \label{fig:valcurves}
\end{figure}

\section{Training data}
\label{app:data}

\subsection{Label generation}
\label{app:labels}

Every DFT-SMD calculation in this work was generated from a single input template in ORCA~6.0.1. The keyword line is \texttt{!\ wB97M-V def2-TZVPD EnGrad TightSCF DefGrid3 [RKS or UKS] SMD(<solvent>)}. The SCF type follows the OMol25 calculation of the same conformer: \texttt{UKS} wherever OMol25 ran the gas-phase calculation unrestricted, that is, for every $M > 1$ and for the bond-breaking singlets of the \texttt{reactivity}, Transition-1x, and RGD subsets, with \texttt{\%scf guessmix 20 end} added for unrestricted singlets, and \texttt{RKS} otherwise. All labels are vertical labels obtained without geometry optimization. The energy label is $\Delta E = E_\mathrm{SMD} - E_\mathrm{gas}$, where $E_\mathrm{SMD}$ is the final single-point energy of the solvated calculation and includes the CDS term, and the force label is $\Delta\mathbf{F} = -(\nabla E_\mathrm{SMD} - \nabla E_\mathrm{gas})$; both are read from the \texttt{.engrad} files and converted from Hartree and Hartree/Bohr to eV and eV/\AA. For the OMol25 conformers the gas-phase energy and gradient are those stored in OMol25, computed with ORCA~6.0.0 at the same level of theory; for the AIMNet2 anions and the charge-separation augmentation the gas-phase calculation was recomputed with the same keyword line without \texttt{SMD}. The 36 solvents that appear in training are water, acetone, acetonitrile, aniline, benzaldehyde, benzene, bromobenzene, carbon tetrachloride, dichloromethane, chloroform, chlorobenzene, carbon disulfide, cyclohexanone, 1,2-dichloroethane, diiodomethane, 1,4-dioxane, DMF, DMSO, ethanol, diethyl ether, ethyl acetate, $n$-hexadecane, $n$-hexane, iodobenzene, methanol, nitromethane, $N$-methylformamide, 1-octanol, $o$-dichlorobenzene, $n$-pentane, 1-pentanol, 1-propanol, 2,2,2-trifluoroethanol, THF, toluene, and tributyl phosphate.

\subsection{Solute sources and sampling}
\label{app:solutes}

\begin{figure*}[htb]
    \centering
    \includegraphics[width=\textwidth]{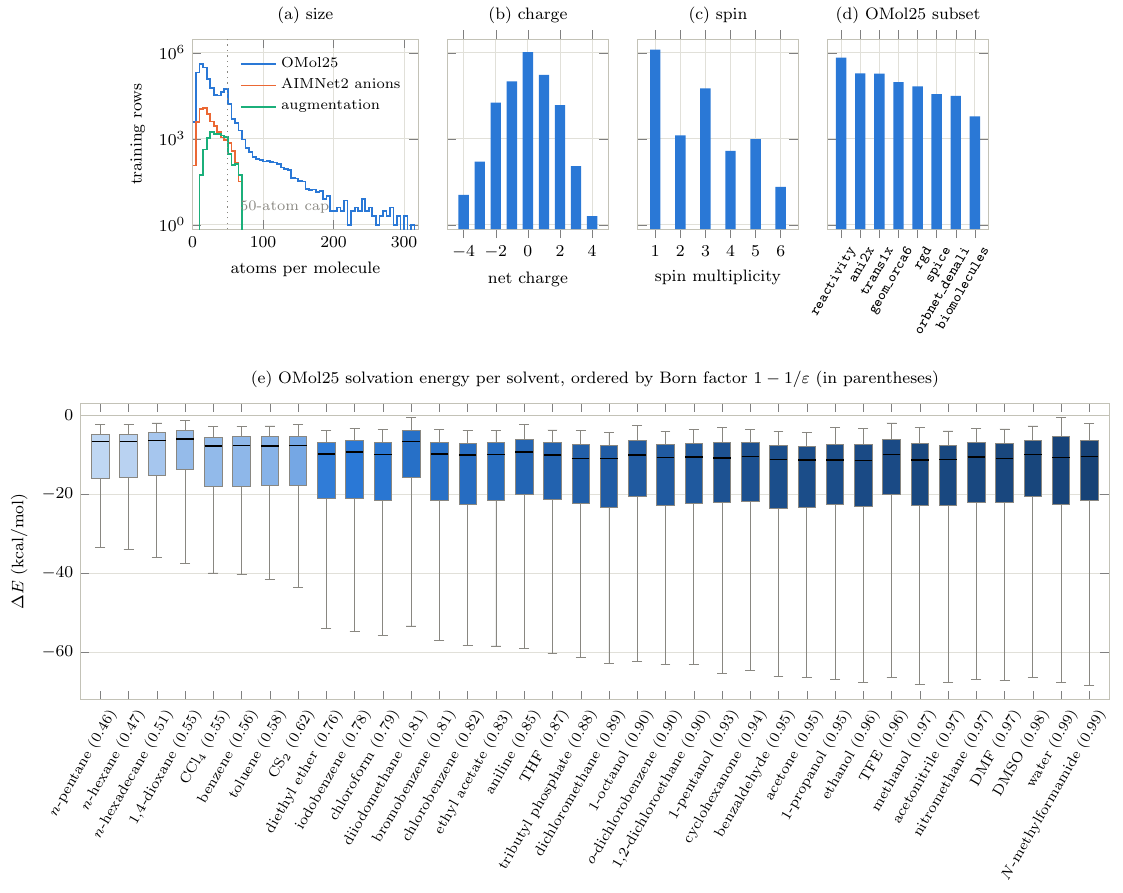}
    \caption{Composition of the training split. (a) Atoms per molecule for the three DFT-labeled blocks, with the 50-atom cap of the non-water solvents marked by the dotted line; (b) net charge; (c) spin multiplicity; (d) OMol25 subset; (e) solvation energy of the OMol25 labels in each solvent, ordered by the Born factor $1 - 1/\varepsilon$ (in parentheses), with boxes spanning the interquartile range and whiskers the 5th to 95th percentiles.}
    \label{fig:distributions}
\end{figure*}

The majority of the solute geometries were drawn from the 4M training split of OMol25. The metal-complex and electrolyte subsets and the solvated-cluster entries of the biomolecule subset were excluded. For all solvents other than water, no molecules with more than 50 atoms were included in the training set. The split is stored as 80 shards whose order is unrelated to subset, and each solvent was assigned a set of shards and labeled every eligible conformer in them. We generated water labels from ten shards with no size cap and the other solvent labels were generated from one or two shards with the 50-atom cap, so that water has about five times as many labels as a typical solvent and every solvent samples a similar mixture of subsets, charges, and spin states (\cref{tab:solvent-counts}). \Cref{fig:distributions} summarizes the composition of the training split.

Geometries and charges of the anion set were taken from the AIMNet2 $\omega$B97M-D3 data set,\cite{anstine202504} restricted to anions with $q = -1$ or $-2$ and sampled in proportion to the size distribution of that data set over 2--69 atoms: 12,783 conformers, giving 46,166 training labels. The four solvents for which these labels were generated, water, DMSO, acetonitrile, and methanol, are those with anion coverage in the experimental benchmarks, and their anion solvation energies fall in a range that the OMol25 data set covers sparsely.

\subsection{Charge-separation augmentation}
\label{app:augmentation}

The charge-separation augmentation species were built by pairing one of four cationic groups, ammonium, guanidinium, $N$-alkylimidazolium, and $N$-alkylpyridinium, with one of four anionic groups, carboxylate, sulfonate, 4-alkylphenolate, and phosphate monoester, through one of four linker classes: alkyl chains $(\mathrm{CH_2})_k$, oligoether chains $\mathrm{CH_2CH_2(OCH_2CH_2)}_k$, aryl--alkyl linkers with a 1,4-phenylene ring at the midpoint, and partially unsaturated linkers with a C=C double bond at the midpoint. The chain lengths were chosen so that the extended conformers span charge separations of roughly 4--22~\AA{} within 50 atoms, and 41 further pairs above 50 atoms extend the separation to 30~\AA. In addition to these zwitterions, two more types of molecules were included in the dataset: 50 charged controls (mono-cations, mono-anions, and same-sign dications and dianions on alkyl or oligoether chains) and 54 contact ion pairs, 3 of them held out. The held-out set, released as part of the test split, comprises the pyridinium--phenolate combination on every linker and every pair on the partially unsaturated linker, 1,197 labels in all. The $\omega$-amino acids of \cref{sec:longrange} were excluded from the library.

\subsection{Splits and training set composition}
\label{app:splits}

Each conformer or molecule belongs to exactly one split across all solvents; the OMol25 split was fixed when the first labels were generated and reused for every later solvent. The OMol25 conformers were split 95/2.5/2.5 into training, validation, and test, stratified by OMol25 subset; the AIMNet2 anions were split 90/5/5 at the molecule level; and the charge-separation augmentation was split 97.5/2.5 at the molecule level into training and validation, with the held-out species of \cref{app:augmentation} as its test split. \Cref{tab:solvent-counts} gives the training labels per solvent.

\begin{table}[htb]
    \centering
    \begin{tabular}{lrlr}
    \hline\hline
    \noalign{\vskip 0.3em}
    Solvent & Labels & Solvent & Labels \\[0.3em]
    \hline
    \noalign{\vskip 0.3em}
    \multicolumn{4}{l}{\textit{OMol25 labels}} \\[0.15em]
    acetone & 34,089 & 1,4-dioxane & 34,241 \\
    acetonitrile & 34,249 & ethanol & 34,042 \\
    aniline & 33,804 & ethyl acetate & 33,784 \\
    benzaldehyde & 34,289 & $n$-hexadecane & 34,303 \\
    benzene & 34,341 & $n$-hexane & 34,317 \\
    bromobenzene & 34,142 & iodobenzene & 34,098 \\
    carbon disulfide & 34,123 & methanol & 33,971 \\
    carbon tetrachloride & 34,308 & $N$-methylformamide & 17,221 \\
    chlorobenzene & 33,856 & nitromethane & 33,903 \\
    chloroform & 34,031 & 1-octanol & 34,107 \\
    cyclohexanone & 34,362 & $n$-pentane & 17,119 \\
    $o$-dichlorobenzene & 34,181 & 1-pentanol & 34,340 \\
    1,2-dichloroethane & 34,404 & 1-propanol & 34,090 \\
    dichloromethane & 33,976 & tetrahydrofuran & 34,260 \\
    diethyl ether & 34,237 & toluene & 33,990 \\
    diiodomethane & 34,109 & tributyl phosphate & 16,927 \\
    dimethyl sulfoxide & 34,093 & 2,2,2-trifluoroethanol & 16,962 \\
    $N$,$N$-dimethylformamide & 34,034 & water & 173,285 \\
    \noalign{\vskip 0.3em}
    \hline
    \noalign{\vskip 0.3em}
    \multicolumn{4}{l}{\textit{AIMNet2 anion labels}} \\[0.15em]
    acetonitrile & 11,544 & dimethyl sulfoxide & 11,537 \\
    methanol & 11,542 & water & 11,543 \\
    \noalign{\vskip 0.3em}
    \hline
    \noalign{\vskip 0.3em}
    \multicolumn{4}{l}{\textit{Charge-separation augmentation labels}} \\[0.15em]
    water & 2,464 & tetrahydrofuran & 2,282 \\
    chloroform & 2,286 & $n$-hexane & 2,284 \\
    \noalign{\vskip 0.3em}
    \hline
    \noalign{\vskip 0.3em}
    \multicolumn{3}{l}{OMol25 geometries, 36 solvents} & 1,299,588 \\
    \multicolumn{3}{l}{AIMNet2 anion geometries, 4 solvents} & 46,166 \\
    \multicolumn{3}{l}{Charge-separation augmentation, 4 solvents} & 9,316 \\
    \multicolumn{3}{l}{Experimental monatomic species, 49 solvents} & 211 \\[0.3em]
    \hline\hline
    \end{tabular}
    \caption{Number of training labels per solvent for the OMol25 solute geometries,
    the AIMNet2 anion geometries, and the charge-separation augmentation (upper to
    lower blocks), all with DFT-SMD solvation energies and forces, followed by the totals of the four training-set components. }
    \label{tab:solvent-counts}
\end{table}

\section{Protocol for Calculations}
\label{app:bench}

All calculations in \cref{sec:results} that required relaxed structures followed the same protocol. The gas-phase potential was UMA-S-1.2 and the solvated potential was UMA-S-1.2 plus AnySolv, except in \cref{tab:menshutkin-bases} and \cref{tab:experiment-bases}, where UMA-S-1.2 was replaced by the listed gas-phase model with the protocol otherwise unchanged. Both were evaluated through ASE\@. Minima were relaxed with the BFGS optimizer to a maximum force of 0.02~eV/\AA{} for the comparison with experiment (\cref{sec:experiment}) and 0.01~eV/\AA{} for SAMPL6 and the Menshutkin reaction. Transition states were optimized with Sella\cite{hermes202211} to a maximum force of 0.01~eV/\AA{}. Each optimized structure was checked with a vibrational calculation. A transition state was accepted only if it had exactly one imaginary frequency larger than 100~cm$^{-1}$ in magnitude, and minima were checked for imaginary frequencies larger than 50~cm$^{-1}$. In cases where imaginary frequencies were found at a minimum, the structure was displaced along the imaginary mode and re-optimized up to four times, after which imaginary frequencies up to 100~cm$^{-1}$ were tolerated. Free energies were obtained from harmonic frequencies computed by finite differences of the model forces at each optimized structure at $T = 298.15$~K. For SAMPL6 and the Menshutkin reaction, where free-energy differences between conformers and between stationary points are the quantities of interest, the quasi-harmonic approximation was used, in which vibrational frequencies below 50~cm$^{-1}$ are raised to 50~cm$^{-1}$.\cite{ribeiro201112}

\begin{table}[htb]
    \centering
    \footnotesize
    \setlength{\tabcolsep}{3pt}%
    \begin{tabular}{lrrrrr}
    \hline\hline
    \noalign{\vskip 0.3em}
    & & \multicolumn{2}{c}{UMA-S-1.2} & \multicolumn{2}{c}{MACE-POLAR-1-M} \\
    Solutes & $n$ & $\Delta E_\mathrm{opt}$ & $\Delta G_\mathrm{solv}$ & $\Delta E_\mathrm{opt}$ & $\Delta G_\mathrm{solv}$ \\[0.3em]
    \hline
    \noalign{\vskip 0.3em}
    Solv@TUM neutral & 4,510 & 0.89 ($-$0.24) & 0.99 ($-$0.48) & 0.90 ($-$0.24) & 1.00 ($-$0.49) \\
    FlexiSol neutral &   529 & 2.46 ($+$1.03) & 2.47 ($+$0.52) & 2.46 ($+$1.02) & 2.48 ($+$0.53) \\
    MNSol neutral    & 2,530 & 0.84 ($-$0.35) & 0.95 ($-$0.52) & 0.84 ($-$0.35) & 0.96 ($-$0.52) \\
    MNSol cations    &   132 & 4.60 ($+$3.64) & 4.56 ($+$3.64) & 4.59 ($+$3.64) & 4.57 ($+$3.67) \\
    MNSol anions     &   222 & 5.08 ($+$2.40) & 5.17 ($+$2.60) & 5.06 ($+$2.38) & 5.22 ($+$2.60) \\[0.3em]
    \hline\hline
    \end{tabular}
    \caption{MAE and, in parentheses, mean signed error (model minus experiment), both in kcal/mol, of the MoE model with UMA-S-1.2 or MACE-POLAR-1-M as the gas-phase potential.}
    \label{tab:experiment-bases}
\end{table}

For the S$_\mathrm{N}$Ar survey (\cref{sec:snar}), minima were converged to a maximum force of 0.01~eV/\AA{}, and transition states were optimized with Sella to the same threshold. Transition-state searches were started from the geometries of Kwan et al.\ where available.\cite{kwan201809} A saddle point was accepted as a transition state if it had exactly one imaginary frequency larger than 100~cm$^{-1}$ in magnitude and the corresponding normal mode was predominantly C--X and C--Y stretching, where X is the leaving group and Y the nucleophile. Each transition state was then connected to its minima by an intrinsic reaction coordinate calculation in Sella.

For the molecular dynamics of alanine dipeptide (\cref{sec:alanine_md}), a starting structure was embedded with RDKit\cite{rdkit} and relaxed with UMA-S-1.2. For each phase (gas phase or solvent), one structure of each stable conformer was relaxed on that phase's potential with $\phi$ and $\psi$ held fixed, to obtain the initial structures for the eight metadynamics walkers of that phase. We tested energy conservation in water, starting from the C7 conformer, with 10~ps of Langevin dynamics at 298.15~K followed by 50~ps of microcanonical velocity Verlet dynamics, similar to the test in the original eSEN paper.\cite{fu202502} The total energy stayed within 0.22~meV/atom of its initial value at a 0.5~fs time step and within 1.3~meV/atom at a 1.0~fs time step, the same as with UMA-S-1.2 alone (0.21 and 1.3~meV/atom), with linear drifts below 0.002~meV/atom per ps in all four runs. Well-tempered metadynamics on $\phi$ and $\psi$ was run with PLUMED\cite{tribello201402} under a Langevin thermostat at 298.15~K (friction 1~ps$^{-1}$), with Gaussians of width 0.35~rad and initial height 1.2~kJ/mol deposited every 500 steps, a bias factor of 6, and eight walkers per phase sharing a single bias\cite{raiteri200603} for 250~ps each. The trajectories were reweighted to the unbiased ensemble\cite{tiwary201501} after discarding the first 20\% of each walker's trajectory. Each frame was assigned to one of five basins: $\alpha_\mathrm{R}$ ($\phi < 0$, $-120^\circ < \psi < 50^\circ$), C5 ($\phi < -120^\circ$ outside the $\alpha_\mathrm{R}$ range of $\psi$), C7 and PPII, which share the remaining region with $\phi < 0$ and were separated by the intramolecular C=O$\cdots$H--N distance (below 2.5~\AA{} for C7), and the region $\phi > 0$. The free energies were converged to within 0.1~kcal/mol, as measured by the change over the last quarter of each run in the free-energy differences between $\phi > 0$ and $\phi < 0$ and between C7 and the combined PPII and $\alpha_\mathrm{R}$ basins. On one A100 MIG instance (3g.20gb), a single walker ran at 84~ms per step. The coupling constant $^3J(\mathrm{H^N,H^\alpha})$ was computed for every frame from the Karplus relation $^3J = A\cos^2\theta + B\cos\theta + C$ with $\theta = \phi - 60^\circ$ and averaged with the metadynamics weights. The values given in \cref{sec:alanine_md} are the means of those obtained with the parameters of Vuister and Bax ($A = 6.51$, $B = -1.76$, $C = 1.60$~Hz)\cite{vuister199308} and of Hu and Bax ($A = 7.09$, $B = -1.42$, $C = 1.55$~Hz),\cite{hu199707} which differ by less than 0.1~Hz. For the explicit-solvent comparison, we applied the same basin assignment and Karplus average, with equal weights, to the $10^6$ solute configurations (one every 1~ps) in the \texttt{ISSNet\_ala2\_example} data set provided by Chen et al.\cite{chen202108}

\section{Compact model results}
\label{app:compact}

The two models are compared in \cref{tab:compact} on the charge-separated zwitterions (\cref{sec:longrange}) and the SAMPL6 partition coefficients (\cref{sec:logp}).

\begin{table}[htb]
    \centering
    \footnotesize
    \setlength{\tabcolsep}{8pt}%
    \begin{tabular}{lcc}
    \hline\hline
    \noalign{\vskip 0.3em}
    & MoE & compact \\[0.3em]
    \hline
    \noalign{\vskip 0.3em}
    \multicolumn{3}{l}{\textit{SAMPL6: RMSE of $\log_{10} P$ against experiment}} \\[0.15em]
    Boltzmann ensemble & 0.51 & 0.51 \\
    lowest-free-energy conformer & 0.53 & 0.54 \\
    \noalign{\vskip 0.3em}
    \hline
    \noalign{\vskip 0.3em}
    \multicolumn{3}{l}{\textit{Zwitterions: error against DFT-SMD (kcal/mol)}} \\[0.15em]
    extended, MAE / max & 0.31 / 1.14 & 0.25 / 1.27 \\
    cyclic, MAE / max & 0.18 / 1.39 & 0.21 / 1.54 \\[0.3em]
    \hline\hline
    \end{tabular}
    \caption{Comparison of the MoE and compact models on the charge-separated zwitterions of \cref{sec:longrange} and the SAMPL6 partition coefficients of \cref{sec:logp}.}
    \label{tab:compact}
\end{table}

\end{document}